\documentclass[aps,prb,floatfix,amsmath,amssymb,twocolumn,eqsecnum,groupedaddress,showpacs]{revtex4}
\usepackage[dvips]{graphicx}
\usepackage{dcolumn}
\usepackage{bm}
\begin{document}
\title{Theory of quantum impurities in spin liquids}

\author{Alexei Kolezhuk}
\affiliation{Department of Physics, Harvard University, Cambridge MA
02138}

\author{Subir Sachdev}
\affiliation{Department of Physics, Harvard University, Cambridge MA
02138}

\author{Rudro R. Biswas}
\affiliation{Department of Physics, Harvard University, Cambridge MA
02138}
\author{Peiqiu Chen}
\affiliation{Department of Physics, Harvard University, Cambridge MA
02138}

\begin{abstract}
We describe spin correlations in the vicinity of a generalized
impurity in a wide class of fractionalized spin liquid states. We
argue that the primary characterization of the impurity is its
electric charge under the gauge field describing singlet excitations
in the spin liquid. We focus on two gapless U(1) spin liquids
described by 2+1 dimensional conformal field theories (CFT): the
staggered flux (sF) spin liquid, and the deconfined critical point
between the N\'eel and valence-bond-solid (VBS) states. In these
cases, the electric charge is argued to be an exactly marginal
perturbation of the CFT. Consequently, the impurity susceptibility
has a $1/T$ temperature dependence, with an anomalous Curie constant
which is a universal number associated with the CFT. One unexpected
feature of the CFT of the sF state is that an applied magnetic field
does not induce any staggered spin polarization in the vicinity of
the impurity (while such a staggered magnetization is present for
the N\'eel-VBS case). These results differ significantly from
earlier theories of vacancies in the sF state, and we explicitly
demonstrate how our gauge theory corrects these works. We discuss
implications of our results for the cuprate superconductors, organic
Mott insulators, and graphene.
\end{abstract}




%

\date{\today}

\maketitle

\section{Introduction}
\label{sec:intro}

The response of a strongly interacting electronic system to
impurities has long been a fruitful way of experimentally and
theoretically elucidating the subtle correlations in its many-body
ground state wavefunction. The most prominent example is the Kondo
effect, which describes the interplay between a variety of
impurities with a spin and/or `flavor' degree of freedom and a
system of free fermions with either a finite
\cite{hewson,nozieres,cmv} or vanishing \cite{vojta} density of
states at the Fermi energy.

More recently, the impurity responses of a variety of
`non-Fermi-liquid' bulk states have been studied.
\cite{kane,affleck,science,sengupta,qimplong,stv,japan,si1,zarand,si2,troyer,qimp2,sandvik,florens}
The $S=1/2$ antiferromagnetic spin chain generically has a critical
ground state, and displays interesting universal characteristics in
its response to impurities or boundaries \cite{affleck}.
Universality was also found in the general theory
\cite{science,qimplong,qimp2} of impurities in `dimerized' quantum
antiferromagnets in spatial dimensions $d \geq 2$ near a quantum
critical point between a N\'eel state and a confining spin gap
state. Such `dimerized' antiferromagnets have an even number of
$S=1/2$ spins per unit cell, and consequently their bulk quantum
criticality is described within the conventional
Landau-Ginzburg-Wilson (LGW) framework of a fluctuating N\'eel order
parameter.\cite{senthil1,senthil2,senthil3} Away from the impurity,
such systems only have excitations which carry integer spin.

It is the purpose of this paper to extend the above theory
\cite{science,qimplong,qimp2} to fractionalized `spin liquid' states
in spatial dimensions $d \geq 2$ with neutral $S=1/2$ excitations
(`spinons') in the bulk. Such spinon excitations carry gauge charges
associated with an `emergent' gauge force (distinct from the
electromagnetic forces), typically with the gauge group
\cite{rs,wen1} $Z_2$ or U(1), and we will argue shortly that such
gauge forces play a key role in the response of spin liquid states
to impurities. Earlier analyses\cite{kk,nn,nl,pepin,ziqiang} of the
influence of impurities in the U(1) `staggered-flux' spin liquid
ignored the crucial gauge forces; we will comment in detail on the
relationship of our results to these works in Section~\ref{sec:kk}.
An analysis of impurities in a $Z_2$ spin liquid state was presented
recently by Florens {\em et al.}, \cite{florens} but for a
particular situation in which a spin moment was strongly localized
on an impurity and gauge forces could be safely neglected; we will
comment further on their work in Section~\ref{sec:z2}.

There are a number of experimental motivations for our work. A large
number of experiments have studied Zn and Ni impurities in the
cuprates,\cite{bobroff,ouazi,yazdani,pan} and much useful
information has been obtained on the spatial and temperature
dependence of the induced moments around the impurity. It would
clearly be useful to compare these results with the corresponding
predictions for different spin liquid states, and for states
proximate to quantum critical points. We will show here that there
are significant differences in the experimental signatures of the
different candidates, and this should eventually allow clear
discrimination by a comparison to experimental results. A second
motivation comes from a recent nuclear magnetic resonance (NMR)
study \cite{shimizu} of the $S=1/2$ triangular lattice organic Mott
insulator $\kappa$-(ET)$_2$Cu$_2$(CN)$_3$, which possibly has a
non-magnetic, spin singlet ground state. The NMR signal shows
significant inhomogeneous broadening, indicative of local fields
nucleated around impurities. Our theoretical predictions here for
Knight shift around impurities should also assist here in selecting
among the candidate ground states.

An important observations is that in situations with deconfinement
in the bulk, the bulk spinons are readily available to screen any
moments associated with an impurity atom (as has also been noted by
Florens {\em et al.} \cite{florens}). Moreover, for a non-magnetic
impurity (such as Zn on a Cu site), there is no {\em a priori}
reason for the impurity to acquire a strongly localized moment.
Consequently, it is very useful to consider the case where the
impurity has local net spin $S=0$. Naively, such a situation might
seem quite uninteresting, as there is then no local spin degree of
freedom which can interact non-trivially with the excitations of the
spin liquid. Indeed, in a Fermi liquid, a non-magnetic impurity has
little effect, apart from a local renormalization of Fermi liquid
parameters, and there is no Kondo physics. However, in spin liquids
the impurity can carry an {\em electric gauge charge} $Q$, and the
primary purpose of this paper will be to demonstrate that a
$Q\neq0$, $S=0$ impurity displays rich and universal physics.

We will primarily consider U(1) spin liquids here: then an important
dynamical degree of freedom is a U(1) gauge field $A_\mu$, where
$\mu$ extends over the $d+1$ spacetime directions, including the
imaginary time direction, $\tau$. Our considerations here can also
be easily extended to $Z_2$ spin liquids, and this will be described
later in Section~\ref{sec:z2}. We normalize the $A_\mu$ gauge field
such that the spinons have electric charges $\pm 1$. In the popular
U(1) gauge theories of antiferromagnets on the square lattice (which
we will describe more specifically below), a vacancy will carry a
gauge charge $Q= \pm 1$. Thus, a Zn impurity on the Cu square
lattice site has $Q=\pm 1$. This can be understood by thinking of
the impurity as a localized `holon' in the doped antiferromagnet,
which also carries such gauge charges \cite{nl}.

We will consider theories here with actions of the structure
\begin{equation}
\mathcal{S}= \mathcal{S}_b+\mathcal{S}_{\rm imp},
\end{equation}
where $\mathcal{S}_b$ is the bulk action of the spin liquid in the
absence of any impurity, and $\mathcal{S}_{\rm imp}$ represents the
perturbation due to an impurity which we assume is localized near
the origin of spatial coordinates, $x=0$. We will argue that the
dominant term in $\mathcal{S}_{\rm imp}$ is the coupling of the
impurity to the U(1) gauge field:
\begin{equation}
\label{imp1} \mathcal{S}_{\rm imp} = i Q \int d \tau A_\tau (x=0,
\tau).
\end{equation}
We will demonstrate that additional terms in the impurity action are
unimportant or `irrelevant'. The $\mathcal{S}_{\rm imp}$ above can
be regarded as the remnant of the spin Berry phase that
characterized the impurity in the previous theory
\cite{science,qimplong,qimp2} of dimerized antiferromagnets; the
latter Berry phase for a spin $S$ impurity was $iS$ times the area
enclosed by the path mapped on the unit sphere by the time history
of the impurity spin. An explicit reduction in the spinon
formulation of the spin Berry phase to Eq.~(\ref{imp1}) was
presented in Ref.~\onlinecite{uli}.

We will consider a variety of realizations of the bulk spin liquid
$\mathcal{S}_b$ in this paper. Our primary results will be for U(1)
algebraic spin liquids, \cite{rantner} which are described by 2+1
dimensional conformal field theories (CFT) and we specialize our
presentation to these CFT cases in the remainder of this section. An
algebraic spin liquid has gapless spinon excitations which interact
strongly with the $A_\mu$ gauge field. An explicit realization
appears in the deconfined quantum critical point
\cite{senthil1,senthil2,senthil3} between N\'eel and valence bond
solid (VBS) states, in which the spinons are relativistic bosons
described by the ${\rm CP}^{N-1}$ field theory. Another is found in
the `staggered flux' (sF) phase of SU($N$) antiferromagnets, where
the spinons are Dirac fermions
\cite{rantner,hermele1,hermele2,flavio}. In all these cases, the
algebraic spin liquid is described by a 2+1 dimensional conformal
field theory, and our primary purpose here is to describe the
\emph{boundary conformal field theory} that appears in the presence
of $\mathcal{S}_{\rm imp}$.

Our central observation, forming the basis of our results, is that
$\mathcal{S}_{\rm imp}$ in Eq.~(\ref{imp1}) is an \emph{exactly
marginal} perturbation to the bulk conformal field theory. This
non-renormalization is a consequence of U(1) gauge invariance, which
holds both in the bulk and on the impurity. We will verify this
non-renormalization claim in a variety of perturbative analyses of
the conformal field theory. The claim can also be viewed as a
descendant of the non-renormalization of the spin Berry phase term,
found in Ref.~\onlinecite{qimp2}.

The exact marginality of $\mathcal{S}_{\rm imp}$ has immediate
consequences for the response of the system to a uniform applied
magnetic field $H$. The impurity susceptibility, $\chi_{\rm imp}$,
defined as the change in the total bulk susceptibility due to the
presence of the impurity, obeys
\begin{equation}
\label{chiimpscale} \chi_{\rm imp} = \frac{\mathcal{C}}{T}
\end{equation}
at finite temperature $T$ above a conformal ground state; this can
be extended by standard scaling forms (as in
Ref.~\onlinecite{qimplong}) to proximate gapped or ordered phases,
as we will describe in the body of the paper. We set $\hbar=k_B=1$
and absorb a factor of the magneton, $g \mu_B$, in the definition of
the Zeeman field. With this, $\mathcal{C}$ is a dimensionless
universal number, dependent only upon the value of $Q$, and the
universality class of the bulk conformal field theory.

It is remarkable that the response of the impurity has a Curie-like
$T$ dependence, albeit with an anomalous Curie constant
$\mathcal{C}$ (which is likely an irrational number). This anomalous
Curie response appears even though there is no spin moment localized
on the impurity. In contrast, the earlier results for the LGW
quantum critical point presented in Ref.~\onlinecite{qimplong} had
an unscreened moment present and so a Curie response did not appear
as remarkable. Here, it is due to the deformation of a continuum of
bulk excitations by the impurity, and the $1/T$ power-law is a
simple consequence of the fact that $H$ and $T$ both scale as an
energy. Indeed any other external field, coupling to a total
conserved charge, will also have a corresponding universal $1/T$
susceptibility.

A Curie-like response of an impurity in the staggered flux phase was
also noted early on by Khalliulin and collaborators, \cite{kk} and
others.\cite{nn,pepin} However, in their mean-field analysis, they
associated this response with a zero energy `bound state', and hence
argued that $\mathcal{C}=1/4$. As noted above, the actual
interpretation is different: there is a critical continuum of
excitations, and its collective boundary critical response has a
Curie temperature dependence as a consequence of hyperscaling
properties. Consequently, $\mathcal{C}$ does not equal the Curie
constant of a single spin, and is a non-trivial number which is
almost certainly irrational. We will discuss the earlier work more
explicitly in Section~\ref{sec:kk}.

We will also consider the spatial dependence of the response to a
uniform applied field, $H$, in the presence of an impurity, as that
determines the Knight shift in NMR experiments. The uniform
magnetization density induced by the applied field leads to a Knight
shift $H K_u (x)$; at $T$ above a conformal ground state this obeys
(the scaling form is also as in Ref.~\onlinecite{qimplong}):
\begin{equation}
\label{kuscale} K_u (x) = \frac{(T/c)^{d}}{T} \Phi_u ( xT/c)
\end{equation}
where $c$ is the spinon velocity in the bulk (we assume the bulk
theory has dynamic critical exponent $z=1$, and henceforth set
$c=1$), $\Phi_u$ is a universal function, and the Knight shift is
normalized so that
\begin{equation}
\label{K-norm} \int d^d x K_u (x) = \chi_{\rm imp}.
\end{equation}
The function $\Phi_u (y)$ has a power-law singularity as $y
\rightarrow 0$, with the exponent determined by a `boundary scaling
dimension': this will be described in the body of the paper for the
various cases.

In addition to the locally uniform Knight shift, an impurity in the
presence of a uniform applied field also induces a `staggered'
moment which typically oscillates at the wavevector associated with
a proximate magnetically ordered state. This leads to a staggered
Knight shift, $H K_s (x)$, which we will also consider here. Such a
staggered Knight shift does appear for the deconfined critical
theory describing the N\'eel-VBS transition, and it has a spatial
distribution associated with that of the N\'eel state. However, the
response for the U(1) sF spin liquid is dramatically different. One
of our primary results is that for the scaling limit theory of the
U(1) sF spin liquid, an applied magnetic field in the presence of an
impurity induces {\em none\/} of the many competing orders
\cite{hermele2} associated with the spin liquid. Thus there is no
analog of the `staggered' Knight shift. A subdominant induction of
competing orders can arise upon including irrelevant operators
associated with corrections to scaling; the primary response,
however, is just the induction of a ferromagnetic moment, which has
a slowly-varying, space-dependent envelope in the vicinity of the
impurity specified by $K_u (x)$.

Our conclusions above for the impurity response of the U(1) sF spin
liquid differ from the earlier mean-field
theories.\cite{kk,nn,nl,pepin,ziqiang} They found an induced moment
which had a strong oscillation between the two sublattices of the
square lattice. We demonstrate here that this oscillation disappears
in the continuum field theory which accounts for the gauge
fluctuations. We are not aware of any reason why fluctuation
corrections to the mean-field predictions should be considered
small.

The outline of the remainder of the paper is as follows. In
Section~\ref{sec:cpn}, we will consider the ${\rm CP}^{N-1}$ model
field theory which describes the vicinity of the N\'eel-VBS
transition. The U(1) sF spin liquid will then be described in
Section~\ref{sec:sf}. Other spin liquids, not described by CFTs or
by exactly marginal impurity perturbations, will be briefly
discussed in Section~\ref{sec:other}. Finally, experimental
implications and a summary appear in Section~\ref{sec:conc}.

\section{${\rm CP}^{N-1}$ model}
\label{sec:cpn}

The ${\rm CP}^{N-1}$ model describes the deconfined quantum critical
point between the N\'eel and VBS states on the square
lattice.\cite{senthil1,senthil2,senthil3} It is a field theory of
complex scalars $z_\alpha$, $\alpha = 1 \ldots N$, with global
SU($N$) symmetry and a coupling to the U(1) gauge field $A_\mu$.
Near the transition at which global SU($N$) symmetry is broken we
can work with the effective action
\begin{eqnarray}
\label{CPN-action} \mathcal{S}_b = \int d^D y \biggl\{ |(
\partial_\mu &-& i A_\mu) z_\alpha |^2 + s |z_\alpha |^2 +
\frac{u_0}{2} \left(|z_\alpha|^2
\right)^2  \nonumber \\
&+& \frac{1}{2 e_0^2} \left( \epsilon_{\mu\nu\lambda}
\partial_\nu A_\lambda \right)^2 \biggr\},
\end{eqnarray}
where $y = (\tau, \vec{x})$ is a spacetime coordinate, $\mu$ extends
over  $D$ spacetime indices, $D$ is the dimension of spacetime,
related to the spatial dimensionality, $d$, by
\begin{equation}
D=d+1
\end{equation}
We are interested in the phases of the field theory in
Eq.~(\ref{CPN-action}) as a function of the tuning parameter $s$.
For $s \ll 0$ there is a magnetically ordered phase with global
SU($N$) symmetry broken, while for $s \gg 0$ we have a spin-gap
phase with full SU($N$) symmetry. We are especially interested in
the conformally-invariant critical point which separates these
phases. We will begin by reviewing the critical properties of the
bulk action $\mathcal{S}_b$ alone in the following subsection, and
describe impurity effects in the subsequent subsections.

\subsection{Bulk theory}
\label{sec:bulk}

We will restrict our analysis here to the $\epsilon$ expansion where
\begin{equation}
\epsilon=4-D.
\end{equation}
This was carried out by Halperin {\em et al.}\cite{hlm} some time
ago, in a different physical context. This section will merely
recast their results in our notation, using the field-theoretical
formulation.

As noted by Halperin {\em et al.}, a stable fixed point is obtained
in the $\epsilon$ expansion only for sufficiently large values of
$N$. However, it is expected that in the physical dimension of
$\epsilon=1$, the fixed point may well be stable down to the needed
values of $N$. Our purpose here is to understand the basic features
of the second-order critical point, and its response to impurities:
so we will assume that the value of $N$ is large enough to ensure
stability of the fixed point. This will delineate the essential
scaling structure of the impurity response, but is not expected to
be quantitatively accurate.

It is also possible to analyze this model using the $1/N$ expansion,
directly in $D=3$. We choose not to present this here, because the
results are very similar to the $\epsilon$-expansion, and the
methods are closely related to those used for the $1/N_f$ expansion
in Section~\ref{sec:sf}.

The renormalization proceeds by defining renormalized fields and
coupling constants $u$ and $f$ by
\begin{eqnarray}
z_\alpha &=& Z_z^{1/2} z_{R \alpha} \nonumber \\
u_0 &=& \frac{\mu^\epsilon Z_4}{Z_z^2 S_D} u \nonumber \\
e_0^2 &=& \frac{\mu^\epsilon Z_e}{S_D} f
\end{eqnarray}
where
\begin{equation}
S_D = \frac{2}{\Gamma(D/2) (4 \pi)^{D/2}}.
\end{equation}
We work in the Lorentz gauge, in which the $A_\mu$ propagator is
$(\delta_{\mu\nu} - p_\mu p_\nu /p^2)/p^2$. In this gauge the
renormalization constants are
\begin{eqnarray}
Z_4 &=& 1 + \frac{(N+4)u}{\epsilon} + \frac{6f^2}{\epsilon u}
\nonumber \\
Z_e &=& 1 + \frac{N f}{3 \epsilon} \nonumber \\
Z_z &=& 1 + \frac{3 f}{\epsilon} \label{zval}
\end{eqnarray}
{}From this we can determine the anomalous dimension, $\eta_z$ of
the $z_\alpha$; note that this is {\em gauge dependent}
\begin{eqnarray}
\eta_z &=& \mu \frac{d}{d\mu} \ln Z_z \nonumber \\
&=& -3f,
\end{eqnarray}
where this is to be evaluated at the fixed point of the $\beta$
functions in Eq.~(\ref{beta}) below.

The $\beta$ functions are
\begin{eqnarray}
\beta(u) &=& \mu \frac{du}{d \mu} = - \epsilon u + (N+4) u^2 + 6 f^2 - 6 f u \nonumber \\
\beta(f) &=& \mu \frac{df}{d \mu} = -\epsilon f + \frac{N}{3} f^2
\label{beta}
\end{eqnarray}

The renormalization of $|z_\alpha|^2$ determines the critical
exponent $\nu$. This is associated with the renormalization constant
\begin{equation}
Z_2 = 1 + \frac{(N+4)u}{\epsilon}
\end{equation}
In this gauge, there is no contribution to $Z_2$ of order $f$. The
critical exponent $\nu$ is given by
\begin{eqnarray}
\frac{1}{\nu} &=& 2 + \mu \frac{d}{d \mu} \ln \frac{Z_2}{Z_z}
\nonumber \\
&=& 2 - (N+4)u + 3f,
\end{eqnarray}
where, again, this is to be evaluated at the fixed point of the
$\beta$ functions in Eq.~(\ref{beta}).

Finally, we consider the scaling dimensions of gauge-invariant
operators which characterize the observable spin correlations. The
N\'eel order parameter is defined by
\begin{equation}
\phi_a = z_\alpha^{\ast} T^a_{\alpha\beta} z_\beta
\end{equation}
where $T^a$ is a $N \times N$ matrix which is a generator of
SU($N$). We define its anomalous dimension $\eta$ by
\begin{equation}
\mbox{dim}[\phi_a] = (D-2+\eta)/2 \label{dimphi}
\end{equation}
In the Lorentz gauge, $\phi_a$ has no additional renormalization
from $A_\mu$ fluctuations at leading order in $\epsilon$. The
anomalous exponent $\eta$ is then given, to this order, by
\begin{equation}
\eta = D-2 + 2 \eta_z + 2u + \mathcal{O}(\epsilon^2) \label{etaeta}
\end{equation}
Note that the value of $\eta$ is gauge-invariant, while that of
$\eta_z$ is not; the relationship (\ref{etaeta}) holds only in the
Lorentz gauge.

We will also be interested in correlations of the magnetization
density $M_a$. This is defined by the response of the system to a
uniform and staggered magnetic fields $H_{u}$ and $H_{s}$, under
which the action is modified by
\begin{eqnarray}
\label{hcoupling} && |(\partial_\tau - i A_\tau ) z|^{2} \rightarrow
\big\{ (\partial_\tau + i
A_\tau ) z_\alpha^{*} + H_{u}^{a} T^{a*}_{\alpha\beta} z^{*}_\beta \big\} \nonumber\\
&& \qquad \times \big\{ (\partial_\tau - i A_\tau ) z_\alpha -
H_{u}^{a} T^a_{\alpha\gamma} z_\gamma \big\} -H_{s}^{a}\phi_{a}.
\end{eqnarray}
The magnetization density is given by
\begin{equation}
M_a = T \frac{\delta \ln \mathcal{Z}}{\delta H_{ua}}
\end{equation}
where $\mathcal{Z}$ is the partition function. Because $H_{u}$
couples to a conserved `charge', it scales as an energy, and
therefore
\begin{equation}
\mbox{dim}[M_a] = d. \label{dimma}
\end{equation}

\subsection{Impurity exponents}
\label{sec:imp}

We now turn to an analysis of spin correlations in the vicinity of
the impurity. The general method is very similar to that followed in
Ref.~\onlinecite{qimp2}.

We assume here that monopole tunneling events remain irrelevant at
the impurity at the quantum critical point, as they do in the bulk.
\cite{senthil1,senthil2,senthil3} Such monopole tunneling events are
defined only in $d=2$, and so their scaling dimensions are not
easily estimated in the $\epsilon$ expansion. However, the monopoles
are irrelevant at large $N$ in the bulk because their action is
proportional to $N$, and the same reasoning applies also in the
presence of the impurity.

As in Refs.~\onlinecite{qimplong,qimp2}, the staggered and uniform
magnetizations operators ($\phi_a$ and $M_a$) acquire additional
renormalizations as they approach the impurity at $x=0$. Let us
denote the corrections by modifying (\ref{dimphi}) to
\begin{equation}
\mbox{dim}[ \phi_a (x \rightarrow 0) ] = \mbox{dim}[\phi_a] +
\Delta_{\rm imp}^\phi \label{dimphib}
\end{equation}
and similarly for $M_a$. We will compute these additional
renormalizations to order $\epsilon^2$ below, and find that to this
order $\Delta_{\rm imp}^\phi = \Delta_{\rm imp}^M$.

It is also useful to restate the above results in the notation of
Refs.~\onlinecite{qimplong,qimp2}. These we find that the
excitations near the impurity are characterized by a spin operator
$S_a$, where $a$ is the index of SU($N$) generators, and we denoted
its scaling dimension by $\eta'/2$, or, equivalently,
\begin{equation}
\langle S_a (\tau) S_b (0) \rangle \sim
\frac{\delta_{ab}}{|\tau|^{\eta'}} \label{sasb}
\end{equation}
at the quantum critical point. Assuming that $S_a$ is obtained as
$\phi_a$ approaches the impurity, we obtain from (\ref{dimphi}) and
(\ref{dimphib}) that
\begin{equation}
\eta' = D-2+\eta + 2\Delta_{\rm imp}^\phi. \label{etap}
\end{equation}
These scaling relations imply \cite{csy} that at the quantum
critical point the temperature dependence of the NMR relaxation rate
is
\begin{equation}
1/T_1 \sim T^{-1+\eta'},
\end{equation}
near the impurity, while $1/T_1 \sim T^{d-2+\eta}$ in the bulk.
Finally, we note that we can also express the above relations using
the the operator product expansion
\begin{equation}
\label{phiS} \lim_{x \rightarrow 0} \phi_a (x,\tau) \sim \frac{S_a
(\tau) }{|x|^{-\Delta_{\rm imp}^\phi}}.
\end{equation}

It is also useful to consider time-dependent correlations of the
magnetization density, $M_a$, as one approaches the impurity. The
initial guess would be that these acquire the additional anomalous
dimension $\Delta_{\rm imp}^M$. However, we will see in our
computations below that the dominant correlations are instead given
by `mixing' between the $\phi_a$ and $M_a$ operators near the
impurity. As was also found in Ref.~\onlinecite{qimplong}, the
impurity Berry phase allows such mixing, and so the operator product
expansion of $M_a$ near the impurity has a dominant term given by
the same operator $S_a$ introduced above to describe the behavior of
$\phi_a$. This assertion is encapsulated in the operator product
expansion
\begin{equation}
 \label{MS}
\lim_{x \rightarrow 0} M_a (x,\tau) \sim \frac{S_a
(\tau)}{|x|^{d-\eta'/2}}.
\end{equation}
The relations (\ref{sasb}), (\ref{phiS}) and (\ref{MS}) are keys to
our analysis, and will allow us to deduce the structure of spin
correlations near the impurity. Their validity will be supported by
a number of perturbative results we will obtain below.

Now we turn to a determination of the exponents $\Delta_{\rm
imp}^\phi$ and $\Delta_{\rm imp}^M$. These are most easily specified
by determining the additional renormalization factor needed to
cancel poles in $\epsilon$ in correlators of $\phi_a$ and $M_a$ as
one approaches the impurity. The simplest correlator of $\phi_a$
which allows us to achieve this is the vertex operator
\begin{equation}
\langle \phi_a (x=0, \omega =0) z_\alpha (k=0, \omega) z^\ast_\beta
(k=0, \omega) \rangle, \label{vertex}
\end{equation}
where the external $z_\alpha$ legs will be truncated, and similarly
for the $M_a$. Here the external frequency $\omega$ is needed to
control the infra-red singularities.

\begin{figure}[t]
\centering \includegraphics*[width=3.1in]{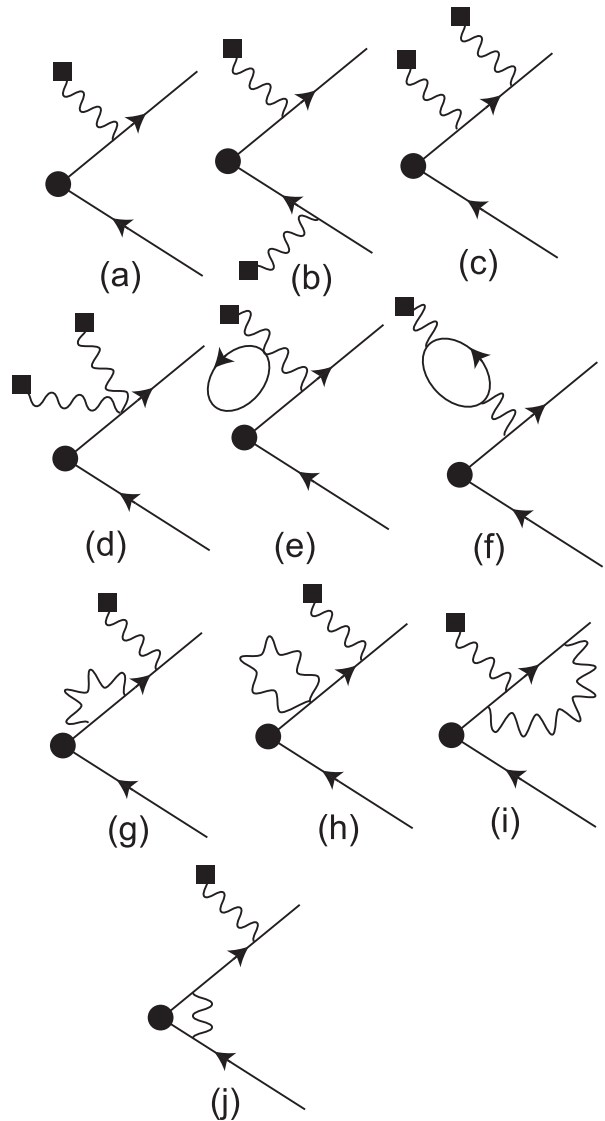} \caption{Feynman
diagrams which contribute to the impurity-dependent renormalization
of vertices such as those in Eq.~(\ref{vertex}). The full line is
the $z_\alpha$ propagator, the wavy line is the $A_\mu$ propagator,
and the filled square is the source term in $\mathcal{S}_{\rm imp}$
(also in Fig.~\ref{fig:vertices-h}e).} \label{fig:etap}
\end{figure}

The leading contribution to (\ref{vertex}), at order $\epsilon$ is
shown in Fig.~\ref{fig:etap}(a) and it renormalized the vertex by
the factor
\begin{equation}
2iQ e_0^2 \omega \int \frac{d^d q}{(2 \pi)^d} \frac{1}{q^2 (q^2 +
\omega^2)} \label{f1a}
\end{equation}
The same expression also applies to the renormalizations of the
vertex associated with $M_a$. The $q$ integral is easily evaluated,
and no pole in $\epsilon$ appears; this is fortunate, because the
expression in (\ref{f1a}) is purely imaginary.

The determination of the $\mathcal{O} (\epsilon^2)$ corrections
involves a more difficult computation. There are a number of graphs
which contribute, shown in Fig.~\ref{fig:etap}.
Fig.~\ref{fig:etap}(b)) is simply the square of
Fig.~\ref{fig:etap}(a), and so contributes no pole in $\epsilon$.
The value of Fig.~\ref{fig:etap}(c) (and its symmetry related
partner) is
\begin{eqnarray}
&& \!\!\! -8Q^{2}e_0^{4}\omega^2
 \int \frac{d^d q}{(2 \pi)^d} \frac{d^d k}{(2 \pi)^d}
\frac{1}{q^2 (q^2 + \omega^2)(k^2 +
\omega^2) (q+k)^2} \nonumber \\
&& \qquad\qquad = \text{no pole in $\epsilon$.}
\end{eqnarray}
The contribution of Fig.~\ref{fig:etap}d (and its symmetry related
partner), for both the $\phi_a$ and $M_a$ vertices is
\begin{eqnarray}
&& 2Q^2 e_0^4  T^a_{\alpha\beta} \int \frac{d^d q}{(2 \pi)^d}
\frac{d^d k}{(2 \pi)^d} \frac{1}{q^2 (k^2 + \omega^2) (q+k)^2}
\nonumber \\
&&\qquad\qquad = Q^2 T^a_{\alpha\beta} \frac{4 \pi^2 f^2}{\epsilon}
+ \ldots
\end{eqnarray}
The diagrams Fig.~\ref{fig:etap}(e-j) are all purely imaginary, so
it must be that the poles in $\epsilon$ in their sum cancel. We have
verified that this is indeed the case. Figs.~\ref{fig:etap}(e,f)
yield a pole in $\epsilon$ which cancels against the second order
contribution from Fig.~\ref{fig:etap}(b) after the charge
renormalization in (\ref{zval}) is accounted for. The self-energy
and vertex corrections in Figs.~\ref{fig:etap}(g-i) cancel against
each other, as usual. And finally, Fig~\ref{fig:etap}(j) represents
an `interference' between two first order contributions, the bulk
renormalization of the vertex and the finite impurity contribution
in Fig.~\ref{fig:etap}(a), and so requires no additional
renormalizations. Collecting these results, and using the fixed
point value of $f$ from (\ref{beta}), we obtain
\begin{equation}
\Delta_{\rm imp}^\phi = \Delta_{\rm imp}^M =  - \frac{72 \pi^2
Q^2}{N^2} \epsilon^2 + \mathcal{O} (\epsilon^3), \label{etap2}
\end{equation}
although $\Delta_{\rm imp}^\phi$ and $\Delta_{\rm imp}^M$ are not
expected to be equal at higher orders.

\begin{figure}[tb]
\includegraphics[width=65mm]{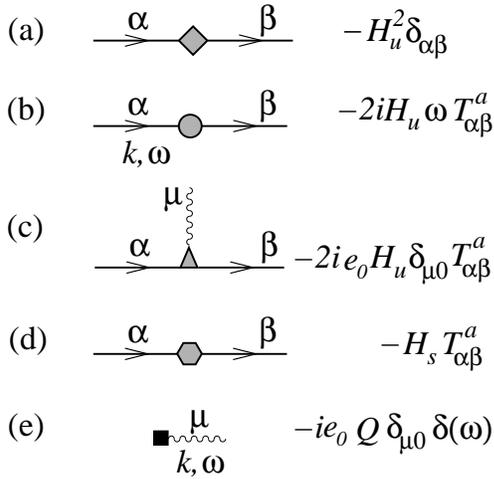}
\caption{\label{fig:vertices-h} Vertices appearing in the
calculation of response to applied field; $H_{u}$ and $H_{s}$ denote
uniform and staggered magnetic fields, respectively. }
\end{figure}

\subsection{Linear response to a uniform applied field}
\label{sec:cp1uni}

We will be interested in the spatially dependent magnetization
response to an external static uniform magnetic field $H_{u}=H$. Let
us start with the calculation of the impurity susceptibility
$\chi_{\rm imp}$, which characterizes the total excess magnetization
arising due to the impurity in response to the applied field.

\subsubsection{Impurity susceptibility}

The susceptibility can be defined as
\begin{equation}
\label{susc-def} \chi_{\rm
imp}=\frac{T}{N^{2}-1}\sum_{a}\frac{\delta^{2} \ln
  \mathcal{Z}}{\delta H_{ua}\delta H_{ua}},
\end{equation}
where the prefactor stems from averaging over the ``spin'' space of
$\rm SU(N)$ generators.
Using the action (\ref{CPN-action}), modified according to
(\ref{hcoupling}), one can
 see that the presence of a uniform magnetic field leads to three new
vertices shown in Fig.\ \ref{fig:vertices-h}(a-c). The
impurity-related diagrams must contain at least one impurity
``source'' vertex shown in Fig.\ \ref{fig:vertices-h}(e).

\begin{figure}[tb]
\includegraphics[width=75mm]{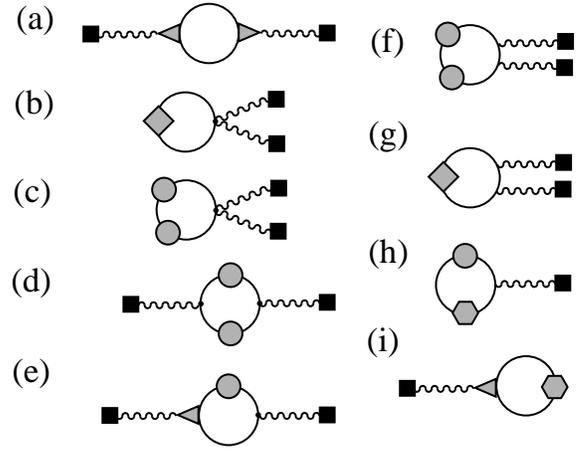}
\caption{\label{fig:diags-h} Feynman diagrams determining the
leading order contribution to the free energy in presence of uniform
(diagrams (a)- (g)) and staggered (diagrams (h), (i)) magnetic
fields.}
\end{figure}

It is easy to show that the leading order correction to the free
energy is given by the diagrams of Fig.\ \ref{fig:diags-h}(a-g).
Notice that the leading contribution is of the second order in the
impurity charge $Q$, since all first-order diagrams happen to be odd
in Matsubara frequency and thus are identically zero. The $A_{\mu}$
propagator enters those diagrams only as $D_{\mu
0}(k,\omega=0)=\delta_{\mu 0}/k^{2}$, and the latter expression is
valid for any choice of gauge
 which ensures that our final results will be gauge-independent.
Summing up the contributions of the diagrams (a-g) of Fig.\
\ref{fig:diags-h}, one obtains
\begin{eqnarray}
\label{chi-tot-src} \chi_{\rm  imp}&=&
2Q^{2}(e_{0}^{2})^{2}\widetilde{S}\int\frac{d^{d}k}{(2\pi)^{d}}
\frac{d^{d}k'}{(2\pi)^{d}}\frac{1}{(k')^{4}}
g(\varepsilon_{k},\varepsilon_{k+k'}),
\nonumber\\
g(x,y)&=&T\sum_{\omega_{n}} \Bigg\{
\frac{2}{(\omega_{n}^{2}+x^{2})(\omega_{n}^{2}+y^{2})}
+\frac{1}{(\omega_{n}^{2}+x^{2})^{2}}\nonumber\\
&-&\frac{4\omega_{n}^{2}}{(\omega_{n}^{2}+x^{2})^{3}}
+\frac{8\omega_{n}^{4}}{(\omega_{n}^{2}+x^{2})^{2}(\omega_{n}^{2}+y^{2})^{2}}\nonumber\\
&-&\frac{16\omega_{n}^{2}}{(\omega_{n}^{2}+x^{2})^{2}(\omega_{n}^{2}+y^{2})}
+\frac{16\omega_{n}^{4}}{(\omega_{n}^{2}+x^{2})^{3}(\omega_{n}^{2}+y^{2})}\nonumber\\
&-&\frac{4\omega_{n}^{2}}{(\omega_{n}^{2}+x^{2})^{2}(\omega_{n}^{2}+y^{2})}
\Bigg\},
\end{eqnarray}
where $\varepsilon_{k}\equiv |k|$, and the sum in $g(x,y)$ runs over
bosonic Matsubara frequencies $\omega_{n}=2\pi n T$. The overall
factor of $\widetilde{S}$  comes from a trace over the components of
the field $z$ and is defined as
\begin{equation}
\label{casimir} \widetilde{S}=\frac{1}{N^{2}-1}\text{tr}
\sum_{a=1}^{N^{2}-1}
(T^{a})^{2}\equiv\frac{\widetilde{\mathcal{C}}}{N^{2}-1},
\end{equation}
where $\widetilde{\mathcal{C}}$ is the eigenvalue of the Casimir
operator of $\rm SU(N)$ which depends on the specific representation
of the group. In the $\rm SU(2)$ case, this factor takes the
familiar form $\widetilde{S}=S(S+1)/3$. It is worth noting that all
the diagrams are symmetric with respect to the exchange of $k$ and
$k+k'$, so only the symmetric part of $g(x,y)$ yields a finite
contribution to the susceptibility.

Obviously, the expression (\ref{chi-tot-src}) contains a singularity
at $k,k'\to 0$. At $T=0$ this singularity is formally unimportant,
since at $T\to 0$ the function $g(x,y)$ becomes fully antisymmetric:
\begin{equation}
\lim_{T\to 0} g(x,y)\to \frac{x-y}{2xy(x+y)^{2}},
\end{equation}
so at $T=0$ the impurity susceptibility is identically zero.
Alternatively, one may notice that the expression under the sum sign
in (\ref{chi-tot-src}) is a full derivative in $\omega_{n}$, which
implies that $\chi_{\rm imp}$ vanishes at $T=0$ when the sum in
$\omega_{n}$ is replaced by an integral.

\begin{figure}[tb]
\includegraphics[width=55mm]{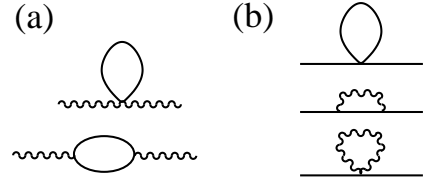}
\caption{\label{fig:tmass} Self-energy diagrams determining the
thermal screening masses (\protect\ref{tmass-src}) for the
``photon'' $A_{\mu}$  (a) and for $z$-boson  (b).}
\end{figure}

At finite $T$, however, the symmetric part
$\widetilde{g}(x,y)=1/2\big( g(x,y) +g(y,x)\big)$ is finite:
\begin{equation}
\widetilde{g}(x,y) =\frac{x(3y^{2}+x^{2})\cosh(y/2T) } {16
xy(x^{2}-y^{2})T^{2}\sinh^{2}(y/2T)} + (x\leftrightarrow y),
\end{equation}
and the infrared singularity becomes dangerous (the integrand in
(\ref{chi-tot-src}) behaves as $k^{-4}(k')^{-4}$ at $k,k'\to 0$).
Such finite-$T$ infrared divergencies in massless theories are
well-known \cite{AndersenStrickland04} and can be cured by
incorporating the thermal ``screening mass'' into free  propagators.
The thermal masses $m_{p}$ and $m_{b}$ (for the ``photon'' $A_{\mu}$
and for the boson $z$, respectively) can be easily found to the
first order in the coupling constants $e_{0}^{2}$ and $u_{0}$ by
computing the temperature-dependent contribution to the self-energy
diagrams shown in Fig.\ \ref{fig:tmass}:
\begin{equation}
\label{tmass-src} m_{p}^{2}=\frac{e_{0}^{2}N}{6}T^{2},\qquad
m_{b}^{2}=\Big\{ \frac{(N+1)u_{0}}{12} + \frac{e_{0}^{2}}{4}
\Big\}T^{2}.
\end{equation}
At the critical point $e_{0}^{2}$ and $u_{0}$ are both proportional
to $\epsilon=4-D$, so that we can write
\begin{equation}
\label{tmass} m_{p,b}^{2}=\lambda_{p,b} T^{2}\epsilon,
\end{equation}
where  $\lambda_{p,b}$ can be found from the beta-functions
(\ref{beta}): one obtains $\lambda_{p}=4\pi^{2}$, and the expression
for $\lambda_{b}$ is rather complicated \cite{Lawrie82} but
simplifies in the large-$N$ limit:
\begin{equation}
\lambda_{p}=4\pi^{2},\quad \lambda_{b}\to \pi^{2}/6, \quad N\to
\infty.
\end{equation}

Taking into account the thermal masses amounts to replacing
$(k')^{2}\mapsto (k')^{2}+m_{p}^{2}$ and $x^{2}\mapsto
x^{2}+m_{b}^{2}$, $y^{2}\mapsto y^{2}+m_{b}^{2}$ in Eq.\
(\ref{chi-tot-src}). It is easy to see that the main contribution to
the integral in (\ref{chi-tot-src})  comes from small momenta
$k,Q\lesssim m_{b,p} \ll T$, so one can replace $\widetilde{g}(x,y)$
by
\begin{eqnarray}
\label{g-zerofreq} \lim_{y,x\to
  0}\widetilde{g}(x,y)&\mapsto& \frac{1/2}{(x^{2}+m_{b}^{2})^{2}} +
  \frac{1/2}{(y^{2}+m_{b}^{2})^{2}} \nonumber\\
 &+& \frac{2}{(x^{2}+m_{b}^{2})(y^{2}+m_{b}^{2}) }.
\end{eqnarray}
This means that the main contribution to the susceptibility comes
from  the zero Matsubara frequency ($\omega_{n}=0$) part of the
expression (\ref{chi-tot-src}) and thus is determined by the first
two diagrams in Fig.\ \ref{fig:diags-h}. Evaluating the momentum
integrals in (\ref{chi-tot-src}) in $d=3$, and substituting
$e_{0}^{2}$  by its fixed-point value $\frac{24\epsilon
\pi^{2}}{N}$, one readily obtains the Curie-like law
(\ref{chiimpscale}) for the impurity susceptibility, with the
anomalous Curie constant $\mathcal{C}$ given by
\begin{equation}
\label{chi-tot} \mathcal{C}=\frac{9\epsilon
  Q^{2}\widetilde{S}\lambda_{p}^{2}}{2\pi^{2} N^{2}} \Big\{
\frac{1}{\lambda_{p}+2\sqrt{\lambda_{p}\lambda_{b}}}
+\frac{1}{4\sqrt{\lambda_{p}\lambda_{b}}} \Big\}.
\end{equation}
It is worthwhile to note that the resulting expression for
$\chi_{\rm imp}$ is proportional to $\epsilon$ and not to
$\epsilon^{2}$ as is apparent from (\ref{chi-tot-src}), because of
the partial cancellation caused by $\epsilon$-dependence of the
thermal masses (\ref{tmass}).

\subsubsection{The Knight shift}

The magnetization response to a uniform magnetic field is
space-dependent due to the presence of an impurity; experimentally,
this can be detected in NMR experiments by measuring the Knight
shift. It is not difficult to generalize the calculation of the
previous subsection to the space-dependent case; the ``uniform''
(slowly varying on the scale of a lattice constant) component of the
Knight shift $K_{u}(x)$ is given by
\begin{equation}
\label{Ku-def} K_{u}(x)=\int d^{d}x' \chi_{u}(x,x'),
\end{equation}
where $\chi_{u}(x,x')$ is a generalization of (\ref{susc-def}),
\begin{equation}
\label{susc-def1}
\chi_{u}(x,x')=\frac{T}{N^{2}-1}\sum_{a}\frac{\delta^{2} \ln
  \mathcal{Z}}{\delta H_{ua}(x)\delta H_{ua}(x')}.
\end{equation}
This uniform Knight shift obeys the scaling form in
Eq.~(\ref{kuscale}). A direct application of Eq.~(\ref{MS}) to this
scaling form implies, as in Ref.~\onlinecite{qimplong} that in the
quantum-critical region
\begin{equation}
K_u (x \rightarrow 0) \sim \frac{1}{T^{1-\eta'/2}}
\frac{1}{|x|^{d-\eta'/2}}. \label{kux0}
\end{equation}

One can see that $\chi_{u}(x,x')$ is determined by the same set of
diagrams of Fig.\ \ref{fig:diags-h}(a-g) but with momenta flowing
into the external vertices. To obtain the Knight shift from
Eq.~(\ref{Ku-def}), it is useful to define the Fourier transform
$\chi_{u}(q)$ by
\begin{equation}
\label{chiuq} K_{u}(x)=\int
\frac{d^{d}q}{(2\pi)^{d}}\chi_{u}(q)e^{iqx},
\end{equation}
and then the diagrams have momentum $q$ flowing out of one external
vertex. Obviously, $\chi_{\rm imp}=\chi_{u}(q=0)$ and the
normalization (\ref{K-norm}) is satisfied. Similarly to the case of
susceptibility considered above, one can show that the Knight shift
is identically zero at $T=0$. At finite temperature the main
contribution to $\chi_{u}(q)$ is given by the diagrams of Fig.\
\ref{fig:diags-h}(a,b) with the Matsubara frequency in each
propagator replaced by the corresponding thermal mass (\ref{tmass}),
yielding
\begin{eqnarray}
\label{chi-u-src} \chi_{u}(q)&\simeq&
2Q^{2}(e_{0}^{2})^{2}\widetilde{S}T\Big\{
I(q,m_{p})I(q,m_{b})  \\
&+&2\int
\frac{d^{d}k}{(2\pi)^{d}}\frac{I(k,m_{b})}{(k^{2}+m_{p}^{2})[(k-q)^{2}+m_{p}^{2}]}
\Big\},\nonumber
\end{eqnarray}
where we have denoted
\[
I(q,m)=\int \frac{d^{d}k}{(2\pi)^{d}}
\frac{1}{(k^{2}+m^{2})[(k+q)^{2}+m^{2}]}.
\]
We will only need the expression for $I(q,m)$ in $d=3$:
\begin{equation}
\label{Iqm} I_{d=3}(q,m)=\frac{1}{4\pi|q|}\arctan\frac{|q|}{2m}.
\end{equation}
Asymptotic expressions for $\chi_{u}(q)$ can be easily obtained. For
small wave vectors $q\ll 2m_{b}$  one recovers the results for the
impurity susceptibility from the previous subsection, $\chi_{u}(q\ll
2m_{b}) \simeq \chi_{\rm imp}$, while for large wave vectors $q\gg
m_{p}$ one has
\begin{equation}
\label{chi-u-res} \chi_{u}(q)\simeq \widetilde{S}T \left(
\frac{12\pi Q
  \epsilon}{N}\right)^{2}
\frac{\ln(q/m_{p})}{q^{2}},\quad q\gg m_{p}.
\end{equation}
At large distances $r\gg m_{p}^{-1}$ the Knight shift decays
exponentially as $e^{-m_{p}r}$ due to the finite photon mass
$m_{p}$. At small distances the Knight shift has the following
behavior:
\begin{equation}
\label{K-u-res} K_{u}(r)\simeq 36\pi\widetilde{S}T \left( \frac{ Q
  \epsilon}{N}\right)^{2} \frac{1}{r}\ln\left(\frac{1}{m_{p}r}
  \right),\quad m_{p}r \ll 1.
\end{equation}
Notice that $m_{p}^{2}\propto \epsilon T^{2}$, so at low temperature
the condition $m_{p}r \ll 1$ will be valid for a wide range of $r$.
Apart from the log correction, (\ref{K-u-res}) is consistent with
the general scaling forms (\ref{kuscale}), (\ref{MS}), and
(\ref{kux0}); the logarithm is likely an artifact of the
integer-valued exponents that appear at low orders in the
computation.

In a similar manner, one can calculate the \emph{staggered Knight
shift}
\begin{equation}
\label{Ks-def} K_{s}(x)=\frac{T}{N^{2}-1}\sum_{a}\frac{\delta^{2}
\ln
  \mathcal{Z}}{\delta H_{ua}\delta H_{sa}(x)}.
\end{equation}
This obeys a scaling form analogous to Eq.~(\ref{kuscale})
\begin{equation}
K_s (x) = \frac{T^{(D-2+\eta)/2}}{T} \Phi_s (Tx)
\end{equation}
where the behavior as $x \rightarrow 0$ analogous to
Eq.~(\ref{kux0}) is
\begin{equation}
K_s (x \rightarrow 0) \sim \frac{1}{T^{1-\eta'/2}}
\frac{1}{|x|^{-\Delta^\phi_{\rm imp}}}. \label{ksx0}
\end{equation}

The staggered Knight shift is computed by the Fourier transform of
the momentum-dependent staggered susceptibility $\chi_{s}(q)$ which
is in the leading order determined by the diagrams of Fig.\
\ref{fig:diags-h}(h,i), with an uncompensated momentum $q$ being
``injected'' in the staggered magnetic field vertex of Fig.\
\ref{fig:vertices-h}(d). Notice that for the staggered
susceptibility the first nonvanishing contribution appears already
in the first order in the impurity charge $Q$. As in the uniform
case, the staggered susceptibility can be proven to be identically
zero at $T=0$, and at finite temperature the result is again
determined by the zero Matsubara frequency term and thus by the
diagram Fig.\ \ref{fig:diags-h}(i), with the zero frequency replaced
in each propagator by the respective thermal mass. This yields
\begin{equation}
\label{chi-s-src} \chi_{s}(q)=\frac{48\epsilon \pi^{2}Q\widetilde{S}
T}{N} \frac{I(q,m_{b})}{q^{2}+m_{p}^{2}} .
\end{equation}
Using (\ref{Iqm}),  one can easily deduce the asymptotic behavior of
the staggered Knight shift  at small distances $r\ll m_{p}^{-1}$:
\begin{equation}
\label{K-s-res} K_{s}(r)=\frac{3\epsilon Q\widetilde{S}T}{N}\ln
\Big(\frac{1}{m_{p}r}
  \Big), \quad m_{p}r \ll 1.
\end{equation}
Again, apart from the logarithmic correction, this result is in
accord with the scaling form suggested by (\ref{phiS}) and
(\ref{ksx0}). It is worthwhile to note that in ${\rm CP}^{N-1}$
model for small $\epsilon=4-D$ and large $N$ the staggered
magnetization response $K_{s}(x)$ to an external magnetic field is
much stronger than the uniform response $K_{u}(x)$.


\subsection{N\'eel ordered phase}
\label{sec:neel}

In the N\'eel phase, one of the components of the $z$ field gets
condensed. Assuming $\langle z_\alpha \rangle \propto \delta_{\alpha
  1}$, we can
parameterize
\begin{equation}
z_\alpha = \left\{ \begin{array}{cc} \left( \sqrt{|s|/u} + \sigma
\right)e^{i \zeta} & ~\mbox{for $\alpha=1$} \\
\psi_\alpha e^{i \zeta} & ~\mbox{for $\alpha > 1$}
\end{array}
\right. \label{zpsi}
\end{equation}
where $\sigma$ is a real field, and $\psi_\alpha$, $\alpha=2\ldots
N$, are $N-1$ complex fields. The overall phase $\zeta$ can be
gauged away, and we can set $\zeta = 1$. Then $A_\mu$ becomes a
massive gauge boson, with the ``mass'' $\sqrt{2 e_0^2 |s|/u}$
acquired by the Higgs mechanism. Inserting (\ref{zpsi}) in
$\mathcal{S}_b + \mathcal{S}_{\rm imp}$ we can obtain a
straightforward perturbative expansion for all physical properties
in powers of $u_0$ and $e_0^2$.

First we describe the behavior of the N\'eel order near the
impurity. We expect this to obey the scaling form
\begin{equation}
\langle \phi_a (x) \rangle = |s|^\beta F_\phi ( x/|s|^\nu )
\end{equation}
where $\beta = (D-2+\eta) \nu/2$. For large $x$, we expect that
$F_\phi$ approaches a constant which characterizes the bulk order,
while for small $x$ near the impurity we may deduce from (\ref{MS})
that
\begin{equation}
\lim_{y \rightarrow 0} F_\phi (y) \sim
\frac{1}{y^{(D-2+\eta-\eta')/2}} \label{limFp}
\end{equation}
Explicitly, computing $\langle \phi_a \rangle$ by the perturbative
expansion defined above, we find that there are no impurity
dependent corrections at first order beyond tree level. So to this
order, $F_\phi (y)$ is independent of $y$, but $y$ dependent terms
do appear at higher orders in $\epsilon$. The spatially independent
response at this order is also consistent with the exponent relation
(\ref{etap}) and (\ref{limFp}).

Turning next to the uniform magnetization, we now expect from the
scaling relations in Section~\ref{sec:imp} that
\begin{equation}
\langle M_a (x) \rangle = |s|^{d\nu} F_m (x/|s|^\nu ) \label{mscale}
\end{equation}
where now the scaling function obeys
\begin{equation}
\lim_{y \rightarrow 0} F_m (y) \sim \frac{1}{y^{d-\eta'/2}}
\label{limFm}
\end{equation}
It is easy to compute $M_a$ to one-loop order, and we find
\begin{equation}
\langle M_a (x) \rangle = \frac{Q T^a_{11} 2 e_0^2 |s|}{u} \int
\frac{d^d q}{(2 \pi)^d} \frac{e^{i q x}}{q^2 + 2 |s|e_0^2 /u}
\label{max}
\end{equation}
This obeys the scaling form in (\ref{mscale}) with a magnetization
that decays exponentially on a scale set by the ``photon'' mass. For
small $x$,  note that to leading order in $\epsilon$, $\langle M_a
(x) \rangle \sim 1/x$, and this is consistent with (\ref{limFm}) and
the value of $\eta'$.

An interesting property of (\ref{max}) is that the total
magnetization near the impurity is given by
\begin{equation}
\int d^d x \langle M_a (x) \rangle = Q T^a_{11}. \label{magtotal}
\end{equation}
As in Ref.~\onlinecite{qimplong}, this relation is expected to be
exact, and the total magnetization near the impurity is therefore
quantized. The quantization is a consequence of the conservation of
total spin. Here it may be viewed as an interesting consequence of
Meissner flux expulsion induced by the condensation of the
$z_\alpha$. In the presence of an applied field $H_a$, the fact that
the $z_\alpha$ condensate is polarized along the $\alpha=1$
direction, the Meissner effect implies from (\ref{hcoupling}) that
$i A_\tau + H_a T^a_{11}$ will fluctuate around zero. So setting
$A_\tau = i H_a T^a_{11}$, we obtain from (\ref{imp1}) a term in the
effective action which is linear in $H_a$, and whose coefficient is
the total magnetization in (\ref{magtotal}).

\subsection{Spin gap phase}
\label{sec:bind}

Finally, we turn to the spin gap phase which appears for
sufficiently large $s$. We will not include `dangerously irrelevant'
monopole effects here,\cite{senthil1,senthil2,senthil3} which are
believed to lead to VBS order and confinement at the longest length
scales. We will restrict our attention to length scales of order
$1/\Delta$, where $\Delta$ is the spin gap. We will argue that in
$d=2$ a spinon is necessarily bound to the impurity, and so the
onset of confinement at longer scales does not have a significant
influence on the behavior of the impurity.

It is straightforward to extend the $\epsilon$ expansion to the spin
gap phase. The spinons experience the Coulomb potential of the
impurity $\sim 1/r^{d-2}$. This potential may form bound state, but
because of the bulk spin gap, these states are above the ground
state. So the spin gap is preserved in the presence of the impurity
and all magnetic response functions vanish exponentially at low
temperature.

We now argue that these conclusions of the $\epsilon$ expansion are
strongly modified in $d=2$. The key point is that the Coulomb
potential $\sim \ln (r)$ in $d=2$, and this increases without bound
as $r \rightarrow \infty$. Consequently the self energy of a naked
impurity with nonzero charge diverges logarithmically with system
size. Therefore, it always pays to create a spinon above the spin
gap, and bind it to the impurity, and neutralize its charge. The
scale at which this occurs can be estimated using the $1/N$
expansion.\cite{gm} In this expansion, the Coulomb interaction is
screened by the vacuum polarization of spin-anti-spinon pairs to a
value of order $(Q \Delta/N) \ln (R \Delta)$. The spinons obey a
Schr\"o dinger equation in this potential with a mass $\Delta$, and
from this we obtain an estimate for the size of the bound state of
order $R \sim (N/Q)^{1/2} / \Delta$. At length scales larger than
this $R$, the impurity will just appear as an isolated spin moment,
and will therefore contribute the ordinary Curie susceptibility
$=1/(4 T)$ (for a $Q=\pm 1$ impurity).

Thus the impurity simply behaves as a free local moment in the spin
gap phase. Note that we did not have to appeal to confinement
physics to reach this conclusion; confinement, and VBS order,
appears at the parametrically larger scale\cite{gm} of order $a
(a/\Delta)^{\tilde{\alpha} N}$, where $a$ is the short-distance
cutoff.

\section{Staggered flux spin liquid}
\label{sec:sf}

This section will examine the response of a second algebraic spin
liquid to impurities: the staggered flux spin liquid with fermionic
spinon excitations. \cite{rantner,hermele1,hermele2,flavio} We will
see that the scaling structure is rather similar to that of the
critical ${\rm CP}^{N-1}$ model examined in Section~\ref{sec:cpn},
although there will be some crucial differences in some physical
properties.

The low energy excitations of the spin liquid are described by $N_f$
flavors of 2-component Dirac fermions, $\Psi$, coupled to the U(1)
gauge field $A_\mu$ with the action
\begin{equation}
\mathcal{S}_b = \int d^3 y \overline{\Psi} \left[- i \gamma^\mu
(\partial_\mu + i A_\mu) \right] \Psi. \label{diracaction}
\end{equation}
As before, $y = (\tau, \vec{x})$ is the spacetime coordinate, and
$\mu$ extends over the $D=3$ spacetime indices. The Dirac matrices
$\gamma^\mu = (\tau^3, \tau^2, - \tau^1)$, where $\tau^\mu$ are the
Pauli matrices, and the field $\overline{\Psi} = i \Psi^\dagger
\tau^3$ --- we follow the notation of Hermele {\em et
  al.}\cite{hermele2}
The number of flavors is $N_f=4$ for the staggered flux state, and
our results are also extended to the so-called $\pi$-flux state,
which has $N_f = 8$. Our analysis will be carried out, as in the
previous works, in a $1/N_f$ expansion. We have not included a bare
Maxwell term for the gauge field in $\mathcal{S}_b$ because it turns
out to be irrelevant at all orders in the $1/N_f$ expansion.

\subsection{Bulk theory}

The structure of the bulk theory has been described in some detail
in Refs.~\onlinecite{rantner,hermele2}, and we will not repeat the
results here. In the large $N_f$ limit, the propagator of the gauge
field in the Lorentz gauge is
\begin{equation}
D_{\mu\nu} (p) = \left(\delta_{\mu\nu} - \frac{p_\mu
p_\nu}{p^2}\right) \frac{16}{N_f p} \label{dmn}
\end{equation}
This propagator arises from the vacuum polarization of the fermions.
Notice that it is suppressed by a power of $1/N_f$, so the $1/N_f$
expansion can be setup as a perturbation theory in the fermion-gauge
field interaction.

A large number of order parameters can be constructed out of fermion
bilinears, and a detailed catalog has been presented.
\cite{hermele2} Among these are the SU$(N_f)$ flavor currents
\begin{equation}
J_\mu^a = - i \overline{\Psi} \gamma^\mu T^a \Psi,
\end{equation}
certain components of which are the magnetization density and
current. Conservation of this current implies the scaling dimension
\begin{equation}
\mbox{dim} [J_\mu^a ] = 2,
\end{equation}
which is the analog of the relation (\ref{dimma}) in $d=2$ (we will
restrict all discussion in this section to $d=2$).

Also considered were the following quantities:
\begin{equation}
N^a = - i \overline{\Psi} T^a \Psi, \qquad M = -i \overline{\Psi}
\Psi,
\end{equation}
which relate to additional order parameters including the N\'{e}el
order (note that $M$ does \emph{not} correspond to the physical
magnetization). Their scaling dimensions have been computed in the
$1/N_f$ expansion:
\begin{equation}
\mbox{dim} [N^a ] = 2 - \frac{64}{3 \pi^2 N_f} +
\mathcal{O}(1/N_f^2)
\end{equation}

\subsection{Relationship to earlier work}
\label{sec:kk}

Before describing the results of our analysis of the impurity in the
U(1) sF phase, it is useful to describe the earlier analyses in
Refs.~\onlinecite{kk,nn,nl,pepin,ziqiang}. They ignored the $A_\mu$
fluctuations, but instead considered a theory of fermionic spinons
$f_{i \alpha}$ on the sites, $i$, of the square lattice; $\alpha$ is
a spin index. These spinons obey $\sum_\alpha \langle f^{\dagger}_{i
\alpha} f_{i \alpha}\rangle = 1$ on every lattice site except at the
impurity $i=0$. Here, they inserted a vacancy (representing {\em
e.g.\/} a Zn impurity) by including a potential term in the
Hamiltonian
\begin{equation}
\mathcal{H}_{\rm imp} = V \sum_\alpha f^\dagger_{0\alpha} f_{0
\alpha} \label{himp}
\end{equation}
and taking the limit $V \rightarrow \infty$ to prohibit any spinons
from residing on the vacancy. The key physical ingredient in these
analyses is the difference in the spinon occupation number between
the impurity and the bulk:
\begin{equation}
\sum_\alpha \langle f^{\dagger}_{i \alpha} f_{i \alpha}\rangle -1 =
-\delta_{i0}. \label{f0i}
\end{equation}

\begin{figure}[tb]
\includegraphics[width=2.5in]{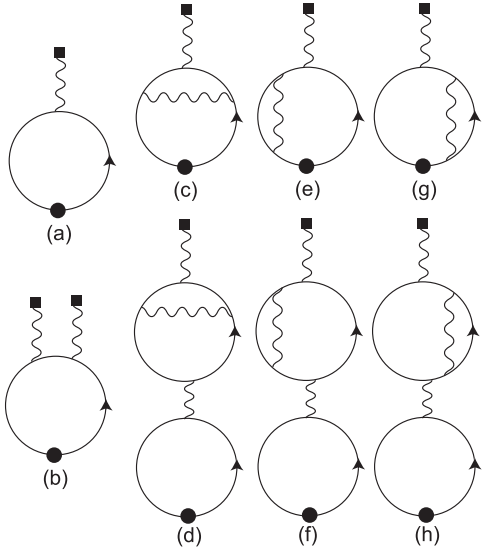}
\caption{Feynman diagrams to order $1/N_f$ for the lhs of
Eq.~(\ref{psi0}). The full line is the Dirac fermion propagator, the
wave line is the photon propagator, the filled square is the
impurity source term in $\mathcal{S}_{\rm imp}$, and the filled
circle is $\Psi^\dagger \Psi$ vertex. As noted in the text, the
diagram in (b) has an odd number of photon vertices, and so vanishes
by Furry's theorem.\cite{furry} Here, and henceforth, we do not show
a number of other diagrams which vanish because of Furry's theorem.}
\label{fig:psi}
\end{figure}

In our approach the analog of the $V=\infty$ limit is obtained by
the functional integral over $A_\tau$.  In the bulk theory, the
continuum field $\Psi$ is defined so that $\langle \Psi^\dagger \Psi
\rangle =0$ in the absence of the impurity, and the continuum analog
of Eq.~(\ref{f0i}) is
\begin{equation}
\langle \Psi^\dagger \Psi (r) \rangle = - Q \delta^2 (r)
\label{psi0}
\end{equation}
where the rhs has a Dirac delta function. The constraint in
Eq.~(\ref{psi0}) is imposed not by adding a potential energy, but by
the functional integral over $A_\tau$, which appears in the
Lagrangian density of $\mathcal{S}_b + \mathcal{S}_{\rm imp}$ (with
$\mathcal{S}_b$ given in Eq.~(\ref{diracaction})) as $i A_\tau \big(
\Psi^\dagger \Psi + Q \delta^2 (r)\big)$. Our treatment of gauge
fluctuations ensures that the constraint on the spinon occupations
is imposed not just on the average, but dynamically on all states
and on all sites. The equality in Eq.~(\ref{psi0}) holds to all
orders in the $1/N_f$ expansion: this follows from the `equation of
motion' for $A_\tau$
\begin{equation}
\left\langle\frac{\delta}{\delta A_\tau} \left( \mathcal{S}_b +
\mathcal{S}_{\rm imp} \right) \right\rangle = 0.
\end{equation}
It is instructive to also test Eq.~(\ref{psi0}) by explicitly
evaluating the lhs of Eq.~(\ref{psi0}) in the $1/N_f$ expansion. The
corresponding Feynman diagrams are shown in Fig.~\ref{fig:psi}. At
leading order, we have the diagram Fig~\ref{fig:psi}(a), which is
easily evaluated to yield the rhs of Eq.~(\ref{psi0}). At order
$1/N_f$, the diagrams shown in Fig.~\ref{fig:psi}(b-g) contribute.
Of these, Fig.~\ref{fig:psi}(b) vanishes by Furry's
theorem.\cite{furry} Of the remaining, it is easy to show that they
cancel in pairs: this requires only the knowledge that the photon
propagator is the inverse of the fermion vacuum polarization bubble.
Thus Figs.~\ref{fig:psi}(c) and (d), (e) and (f), and (g) and (h),
all cancel against each other, and Eq.~(\ref{psi0}) is thus
established to this order. It is not difficult to extend these
arguments to all orders in $1/N_f$. Note that it is possible to
satisfy Eq.~(\ref{psi0}) in a perturbative treatment of gauge
fluctuations, without the need to appeal to bound states: the delta
function at $r=0$ arises from a superposition of the contribution of
many extended states, rather than from a bound state claimed
earlier.\cite{kk,nn,nl}

It is clear that our approach yields a systematic and controlled
treatment of the spinon deficit at the impurity, in contrast to the
earlier ad-hoc mean-field approaches.\cite{kk,nn,nl,pepin,ziqiang}
Our analysis implies that the Curie constant $\mathcal{C}$ is a
non-trivial number, with contributions at all orders in $1/N_f$, and
is not given simply by the Curie constant of a single spin.

The proper analysis of a potential term like that in
Eq.~(\ref{himp}) requires the scaling analysis of perturbations to
the conformal field theory defined by $\mathcal{S}_b +
\mathcal{S}_{\rm imp}$. The action for such a perturbation takes the
form
\begin{equation}
\mathcal{S}^\prime_b = V \int d\tau \Psi^\dagger \Psi (x=0, \tau)
\label{Vpert}
\end{equation}
A simple analysis of scaling dimensions shows that
\begin{equation}
\mbox{dim}[V] = -1 + \mathcal{O}(1/N_f).
\end{equation}
So, potential scattering is an irrelevant perturbation at all orders
in the $1/N_f$ expansion.

A further distinction between our results and the earlier
work\cite{kk,nn,nl,pepin,ziqiang} appears in the response to a
uniform magnetic field. This has significant experimental
consequences, and will be discussed in Section~\ref{sec:sfuni}.

\subsection{Impurity exponents}
\label{sec:imp2}

The analysis is analogous to that carried out in
Section~\ref{sec:imp} for the ${\rm CP}^{N-1}$ model. As there, and
for similar reasons, we neglect the possible influence of monopoles.

Here, we will determine the impurity renormalization to the order
$1/N_f^2$. To this order, we find that all the operators introduced
above, the $J_\mu^a$ and $N^a$ acquire a common correction,
$\Delta_{\rm imp}$, to their bulk scaling dimension. This correction
is given by the single diagram in Fig~\ref{fig:etap}(c), which
(along with its symmetry related partner) evaluates to
\begin{eqnarray}
&& \frac{512 Q^2}{N_f^2} \int \frac{d^2 q}{(2 \pi)^2} \frac{d^2
k}{(2 \pi)^2} \frac{(-i \omega + \vec{q} \cdot \vec{\tau})(-i \omega
+ (\vec{k} + \vec{q}) \cdot \vec{\tau}) }{k q (q^2 +
\omega^2)((q+k)^2 +
\omega^2)} \nonumber \\
&& \qquad= \frac{128 Q^2}{N_f^2 \pi^2} \ln(\Lambda/\omega)
\end{eqnarray}
where $\vec{\tau} = (\tau^1 , \tau^2)$ and $\Lambda$ is an
ultraviolet cutoff; here and henceforth we are using the fermion
lines to represent the propagator $(i \omega + \vec{\tau} \cdot
\vec{q})^{-1}$ from $\Psi$ to $\Psi^\dagger$ (rather then the Dirac
propagator from $\Psi$ to $\overline{\Psi}$). There are a number of
other diagrams, like those shown in Fig.~\ref{fig:etap}, which could
contribute to the vertex renormalization; however they do not
contribute either for reasons as discussed for the ${\rm CP}^{N-1}$
model, or because of Furry's theorem. From this we obtain the
impurity correction to the scaling dimension
\begin{equation}
\label{Delta-imp-SF} \Delta_{\rm imp} = - \frac{128 Q^2}{N_f^2
\pi^2} + \mathcal{O}(1/N_f^3)
\end{equation}
We expect that the higher order corrections will not be the same for
the $J_\mu^a$ and $N^a$.

\begin{figure}[t]
\includegraphics[width=2.2in]{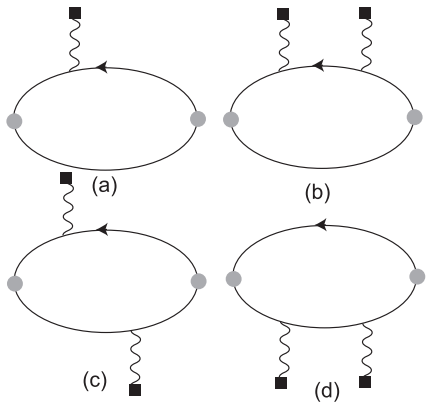}
\caption{Feynman diagrams for the impurity susceptibility and Knight
shifts of the staggered flux spin liquid. The grey circle vertices
depend upon the correlator being evaluated: they equal ({\em i\/})
$T^a \gamma_\tau$ for the ferromagnetic spin density $J_\tau^a$,
({\em ii\/}) $T^a$ for the order parameter $N^a$, and ({\em iii\/})
unity for $M$.} \label{fig:sf}
\end{figure}

\subsection{Linear response to a uniform applied field}
\label{sec:sfuni}

Again, we proceed in analogy to the analysis in
Section~\ref{sec:cp1uni}. We apply a uniform magnetic field $H_a$,
associated with the SU($N_f$) generator $T^a$, which couples
linearly to the conserved total spin density $J_\tau^a$. In
principle, in the presence of the impurity, the linear response to
this applied field can induce time-independent, space-dependent
average values of not only the spin density, $J_\tau^a$, but also of
the other order parameters $M$, and $N^a$. This would be analogous
to the non-zero averages of the uniform and staggered magnetizations
induced by an applied magnetic field on the ${\rm CP}^{N-1}$ model
(which led to the uniform and staggered Knight shifts). However,
here we will find a crucial difference. In the scaling limit of the
algebraic spin liquid represented by Eq.~(\ref{diracaction}), $H_a$
induces {\em only\/} a non-zero $J_\tau^a$, and average values of
all the $N^a$ and $M$ are zero. Thus there is only a uniform Knight
shift, $K_u (x)$, and all the `staggered' Knight shifts, $K_s (x)$,
associated with the many competing orders are zero. The staggered
Knight shift can appear only if some corrections to scaling are
included, associated with irrelevant operators which reduce the
symmetry of the conformal theory to that of the lattice model, and
so can be expected to be weaker than the uniform Knight shift.

The vanishing of $\langle M (x)\rangle $ and $\langle N^a
(x)\rangle$ in the presence of $H_a$ can be established by a careful
consideration of the symmetries of the Dirac fermion theory. First
$\langle M \rangle = 0$, to linear order in $H_a$, simply by
SU($N_f$) symmetry. Establishing the value of $\langle N^a \rangle$
requires more complicated considerations. Let us consider the
leading contribution to $\langle N^a \rangle$, to linear order in
$H_a$, in the $1/N_f$ expansion, represented by the graph in
Fig~\ref{fig:sf}(a). The value of the fermion loop is proportional
to
\begin{eqnarray}
&& T \sum_{\omega_n} \int \frac{d^2 k}{ 4\pi^2} \mbox{Tr}
\bigl[(i\omega_n  + \vec{\tau} \cdot \vec{k})^{-1} \tau^{3}
(i\omega_n + \vec{\tau} \cdot \vec{k})^{-1}
\nonumber \\
&&\qquad \times (i\omega_n + \vec{\tau} \cdot
(\vec{k}+\vec{q}))^{-1}  \bigr]
\end{eqnarray}
Evaluating the trace over the Dirac matrices, we obtain an identical
zero. This can be understood as a consequence of time-reversal
invariance. Both $H_a$ and $N^a$ are odd under the time reversal,
\cite{hermele1} as is the charge of the impurity $Q$. So an
expansion of $\langle N^a \rangle/H_a$ can only involve even powers
of $Q$. Proceeding to order $1/N_f^2$, we obtain diagrams like those
shown in Fig.~\ref{fig:sf}(b-d), which have a pre-factor of $Q^2$,
and so are potentially non-zero. However, these diagrams have a
fermion loop with an odd number of $\gamma^\mu$ vertices, and so
vanish by Furry's theorem. \cite{furry} By a combination of Furry's
theorem and time-reversal invariance we can now easily see that all
terms vanish and so $\langle N^a \rangle = 0$. Because of the $T^a$
matrices in the definitions of $N^a$ and $J^a_\tau$, there must be a
single fermion loop which connects the external vertices. By Furry's
theorem, this loop must have an odd number of photon vertices. All
other fermion loops can only have an even number of photon vertices.
Consequently, there must be an odd number of photon vertices
remaining to connect to the external impurity source term. However,
by time-reversal, there must be an even number of impurity terms;
hence the result.

Our conclusion that $\langle N^a \rangle =0$ is starkly different
from the mean-field prediction \cite{kk,nn,nl,pepin,ziqiang} of an
induced moment which oscillated strongly between the two square
sublattices. Such oscillations can only appear upon including
irrelevant operators.

It remains only to compute the uniform Knight shift $K_u (x)$, or
equivalently, its Fourier transform $\chi_u (q)$ defined in
Eq.~(\ref{chiuq}). The scaling analysis of Section~\ref{sec:imp2}
implies that this Knight shift obeys the scaling form in
Eq.~(\ref{kuscale}), with the $x \rightarrow 0$ behavior given by
\begin{equation}
K_u (x \rightarrow 0) \sim T^{d-1+\Delta_{\rm imp}}
\frac{1}{|x|^{-\Delta_{\rm imp}}}. \label{kux02}
\end{equation}
Here we expect that the $\Delta_{\rm imp}$ exponent is that
associated with $J_\tau^a$, and not (unlike the situation with the
${\rm CP}^{N-1}$ model) that associated with the $N^a$, because
there is no `mixing' between the staggered and uniform
magnetizations near the impurity for the sF spin liquid.

At first order in $1/N_f$, $\chi_u (q)$  is given by the diagram in
Fig~\ref{fig:sf}(a), which vanishes by Furry's theorem. The leading
non-vanishing contribution is at order $1/N_f^2$, and is given by
the diagrams in Fig~\ref{fig:sf}(b-d) (a number of order $1/N_f^2$
diagrams which vanish because of Furry's theorem are not shown). The
sum of these diagrams can we written in the following compact form
\begin{eqnarray}
\label{chiusf} \chi_u (q) &=& Q^2 \widetilde{S} T\sum_{\omega_n}
\int \frac{d^2 k}{4 \pi^2} \frac{d^2 p}{4 \pi^2}
\frac{\partial}{\partial (i\omega_n) } \mbox{Tr} \Biggl[ \left(i
\omega_n + \vec{\tau}
\cdot \vec{k}\right) \nonumber \\
&\times&  \left(i \omega_n + \vec{\tau} \cdot (\vec{k}+\vec{q})
\right) \left(i \omega_n + \vec{\tau} \cdot
(\vec{k}+\vec{p})\right)\Biggr]^{-1} \nonumber \\
&&\qquad \times D_{\tau\tau} (p) D_{\tau\tau} (|\vec{q}-\vec{p}|).
\end{eqnarray}
Here $D_{\tau\tau}$ is the $\tau$ component of the photon
propagator, representing the external lines connecting to the
impurities, and $\widetilde{S}$ is the constant proportional to the
$\rm SU(N)$ Casimir eigenvalue as defined in (\ref{casimir}). Notice
that the expression is a total frequency derivative; this
immediately implies that $\chi_u (q) = 0$ at $T=0$, when the
frequency summation can be converted to an integration. This
vanishing is a consequence of the conservation of total spin.

A non-zero $\chi_u (q)$ is obtained at $T>0$, and we now compute
this. First we need the photon propagator at $T>0$. This is given by
a single fermion loop and at this order we only need the $\tau,\tau$
component at a spatial momentum $\vec{q}$:
\begin{eqnarray}
D_{\tau\tau}^{-1} (q) &=& - N_f T \sum_{\omega_n} \int \frac{d^2
k}{4 \pi^2} \mbox{Tr} \Biggl[ \left(i \omega_n + \vec{\tau} \cdot
\vec{k}\right) \nonumber \\
&&\qquad \times \left(i \omega_n + \vec{\tau} \cdot
(\vec{k}+\vec{q}) \right) \Biggr]^{-1}. \label{dtt1}
\end{eqnarray}
We first combine the denominators in Eq.~(\ref{dtt1}) using the
Feynman parameter $u$, perform the frequency summation, and finally
integrate over $k$. This yields
\begin{equation}
\label{dtt2} D_{\tau\tau}^{-1} (q) = \frac{N_f T}{\pi} \int_0^1 du
\ln \left[ 2 \cosh \left( \frac{q}{2T} \sqrt{u (1-u)} \right)
\right].
\end{equation}
Eq.~(\ref{dtt2}) interpolates between $N_f q/16$ for $T \ll q$ which
agrees with the $T=0$ result in Eq.~(\ref{dmn}), to $(T/\pi)\ln 2$
for $q \ll T$.

\begin{figure}[t]
\includegraphics[width=80mm]{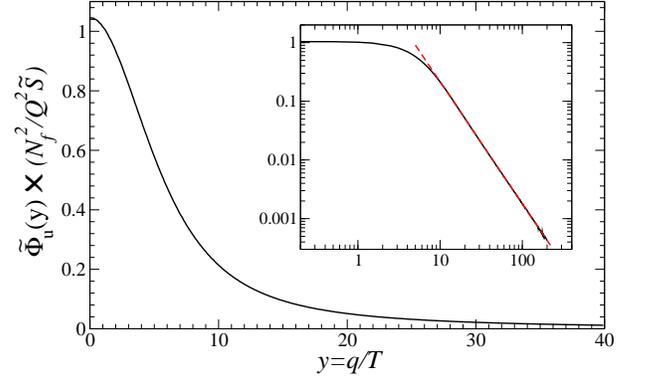}
\caption{ \label{fig:sf-scalfun} Numerically calculated scaling
function $\widetilde{\Phi}_{u}(y)$ defined by
(\protect\ref{dtt2}-\ref{chi-u-q-src2}). The inset shows large-$y$
behavior together with the least square fit (dashed line) to a power
law for $y>40$: the fit yields $\widetilde{\Phi}_{u}(y)\simeq
\mathcal{C}'/y^{\alpha}$ with $\mathcal{C}'=24.7(1)$ and
$\alpha=2.070(5)$. }
\end{figure}

Inserting Eq.~(\ref{dtt2}) into Eq.~(\ref{chiusf}), we conclude that
$\chi_u (q)$ obeys the scaling form
\begin{equation}
\label{chi-u-q-scal} \chi_u (q) = \frac{1}{T} \widetilde{\Phi}_u
(q/T)
\end{equation}
where the scaling function $\widetilde{\Phi}_u$ is the Fourier
transform of the scaling function $\Phi_u$ in Eq.~(\ref{kuscale}).
Further, $\mathcal{C}$, the anomalous Curie constant appearing in
Eq.~(\ref{chiimpscale}), equals $\widetilde{\Phi}_u (0)$. To
determine the scaling function $\widetilde{\Phi}_u$, we need to
evaluate Eq.~(\ref{chiusf}). We combined the Green's functions using
two Feynman parameters $u$, $v$, evaluated the integral over $k$,
differentiated with respect to frequency, and finally evaluated the
summation over $\omega_n$. This yields
\begin{eqnarray}
\label{chi-u-q-src} \chi_u (q) &=& -\frac{Q^2 \widetilde{S}}{\pi}
\int_0^1 du \int_0^{1-u} dv \int \frac{d^2 p}{4
\pi^2} \Bigg\{  f'(\Delta)  \nonumber \\
&+& \bigg[ 2\Delta^{2} - uq^{2}-vp^{2}-\big(1-2(u+v)\big)\vec{p}
\cdot
\vec{q} \bigg]   \nonumber \\
&&\qquad\times \frac{f'' (\Delta)}{4 \Delta} \Bigg\} D_{\tau\tau}
(p) D_{\tau\tau} (|\vec{q}-\vec{p}|)
\end{eqnarray}
where
\begin{equation}
\label{chi-u-q-src1} \Delta^2 = q^2 u(1-u) + p^2 v(1-v) - 2 u v
\vec{p} \cdot \vec{q}
\end{equation}
and $f$ is the Fermi function
\begin{equation}
\label{chi-u-q-src2} f(\varepsilon) = 1/(e^{\varepsilon/T} + 1).
\end{equation}
The remaining integrals have to be evaluated numerically. From such
an evaluation at $q=0$ we found
\begin{equation}
\label{sfc} \mathcal{C} = \frac{1.0460(5) \widetilde{S} Q^2}{N_f^2}
+ \mathcal{O} (1/N_f^3)
\end{equation}
for the universal Curie constant appearing in
Eq.~(\ref{chiimpscale}). The calculated shape of the scaling
function $\widetilde{\Phi}_{u}(y)$ is shown in Fig.\
\ref{fig:sf-scalfun}. From the numerical analysis of its behavior
for large arguments, we deduce that $\widetilde{\Phi}_{u}(y)$ has a
power-law decay $\widetilde{\Phi}_{u}(y)\propto 1/y^{\alpha}$ at
$y\gg1$, where the exponent $\alpha\approx 2$ within the accuracy we
were able to achieve when calculating the four-dimensional integral
in (\ref{chi-u-q-src}). Therefore, we obtain that at small distances
the uniform Knight shift $K_{u}(x)$ tends to a (lattice
cutoff-dependent) constant as $x\to0$, which is consistent, to the
leading order in $1/N_{f}$, with Eq.~(\ref{kux02}) and the fact that
$\Delta_{\rm imp}$ vanishes at this order (see
Eq.~(\ref{Delta-imp-SF})).

\section{Other spin liquids}
\label{sec:other}

This section will discuss the extension of our results to some other
examples of spin liquids. The cases considered below do not share
the high degree of universality found in the critical ${\rm
CP}^{N-1}$ and sF spin liquids above: some aspects depends upon
microscopic details. We will consider the spinon Fermi surface in
Section~\ref{sec:fermi} and $Z_2$ spin liquids in
Section~\ref{sec:z2}

\subsection{Spinon Fermi surface}
\label{sec:fermi}

This spin liquid differs from the sF spin liquid by having a Fermi
surface of gapless spinon excitations, with a non-zero density of
states. The bulk theory is {\em not\/} a CFT, and the bulk action
for the fermionic spinons $f_\alpha$ is
\begin{eqnarray}
\mathcal{S}_b &=& \int d \tau \int \frac{d^d k}{(2 \pi)^d} \Bigl[
f^\dagger_{\alpha} (k,\tau) \Bigl( \frac{\partial}{\partial \tau} +
i A_\tau  \nonumber \\
&& \qquad \qquad\qquad  +\epsilon(\vec{k}-\vec{A}) \Bigr) \Bigr]
f_\alpha (k,\tau)
\end{eqnarray}
Here $\alpha$ is a SU($N$) spinor index, $\epsilon (\vec{k})$ is the
spinon dispersion, and the Fermi surface defined by $\epsilon
(\vec{k}) = 0$ has a finite density of zero energy spinon
excitations. The properties of this spin liquid can be computed by a
combinations of previous methods\cite{aim,km} and the methods
introduced here.

In the absence of the impurity, the spin susceptibility is finite as
$T \rightarrow 0$, equal to the density of states at the Fermi
level. The impurity introduces a `Coulomb' potential for the
spinons. In the small momentum limit, $k \rightarrow 0$, this
interaction is screened to a constant value by the Fermi surface
excitations. Evaluating the response to an applied magnetic field
using a diagram like that in Fig.~\ref{fig:sf}(a), we conclude that
$\chi_{\rm imp} (T \rightarrow 0)$ is finite and proportional to the
energy derivative of the density of states at the Fermi level.

Upon considering the spatial dependence of the Knight shift, again
using the diagram in Fig.~\ref{fig:sf}(a), we find that the dominant
response oscillates at the `$2k_F$' wavevector.\cite{aim} The
transverse gauge interactions strongly enhance the impurity-spinon
vertex, and this leads to a singular response at $2 k_F$ decaying
with a power-law of $x$ determined in Refs.~\onlinecite{aim,km}.
Again, this response is finite as $T \rightarrow 0$, although the
singular portion will have strong corrections at $T>0$.

\subsection{$Z_2$ spin liquid}
\label{sec:z2}

We will describe the $Z_2$ spin liquid accessed from the ${\rm
CP}^{N-1}$ model of Section~\ref{sec:cpn} by condensing a charge-2
Higgs scalar, $\Phi$.\cite{rs,rs2} We add a term like
$|(\partial_\mu - 2 i A_\mu ) \Phi|^2$ to the full action of the
${\rm CP}^{N-1}$ model, and assume we are deep in the phase with
$\langle \Phi \rangle \neq 0$. This condensation gaps out the
$A_\mu$ excitations, and so it is useful to completely integrate out
the $A_\mu$. The result is an effective action for bosonic spinons
$z_\alpha$ which describes a bulk transition from a ordered state
with non-collinear magnetic order to a $Z_2$ spin liquid. This is
proposed as a possible model for the triangular lattice
antiferromagnet.\cite{css,kimsenthil,rmp} The effective action
obtained by this method has the bulk action
\begin{eqnarray}
&& \mathcal{S}_b = \int d^D y \Big\{ |
\partial_\mu z_\alpha |^2 + s |z_\alpha |^2 +
\frac{u_0}{2} \left(|z_\alpha|^2
\right)^2  \nonumber \\
&& \qquad + v_1 \left| z_\alpha^\ast \partial_\mu z_\alpha \right|^2
+ v_2  \left( z_\alpha^\ast \partial_\mu z_\alpha \right)^2 +
\mbox{c.c.} \Big\}, \label{z2}
\end{eqnarray}
and an impurity term which descends from its U(1) charge
\begin{eqnarray}
\mathcal{S}_{\rm imp} &=& \int d\tau \Bigl\{ \lambda_1 z_\alpha^\ast
(x=0,\tau)
\partial_\tau z_\alpha (x=0 ,\tau) \nonumber \\
&&\qquad + \lambda_2 |z_\alpha (x=0,\tau) |^2 \Bigr\}. \label{z2imp}
\end{eqnarray}
Here the coefficient $\lambda_1$ is proportional to the impurity
charge $Q$, but the value of $\lambda_1$ is not quantized. An
initial derivation does not immediately yield the $\lambda_2$ term,
but it is clearly allowed by the symmetries and is evidently more
relevant than the $\lambda_1$ term: it represents a potential
scattering term for the bulk spinons. The critical point of the bulk
theory alone is a CFT, and it is expected that the $v_{1,2}$ are
irrelevant at this critical point, leading to global O($2N$)
symmetry \cite{css,joli}. However, as we will see below, the
perturbations in $\mathcal{S}_{\rm imp}$ are not exactly marginal.

It is useful to compare this theory with that of the recent work of
Florens {\em et al.\/}\cite{florens} Unlike us, they include an
explicit spin degree of freedom at the impurity. In our approach,
such a situation would be appropriate when the couplings
$\lambda_{1,2}$ are large enough to bind a spinon in the ground
state at the impurity. In a $Z_2$ spin liquid, there are no long
range gauge forces (the $Z_2$ gauge forces are short range), unlike
the situation in Section~\ref{sec:bind}, and so it is not required
that the ground state have non-zero spin for a vacancy impurity,
{\em e.g.,\/} Zn on a Cu site. For a magnetic impurity, such as Ni
on a Cu site, a finite spin ground state may be more likely, but is
also not required when the bulk system is in a gapped $Z_2$ spin
liquid. We will restrict our discussion here to the small
$\lambda_{1,2}$ region of (\ref{z2})+(\ref{z2imp}), and this is
equivalent to being in the Kondo-screened phases of Florens {\em et
al.\/}\cite{florens}.

Let us now describe the phases of $\mathcal{S}_b + \mathcal{S}_{\rm
imp}$ as the value of the tuning parameter is scanned.

For $s \gg 0$, we have the gapped spin liquid phase in the bulk. The
bulk spinons will experience a potential from the $\lambda_{1,2}$
terms, and in the attractive case in $d=2$, such a potential always
has a bound state. However, the binding energy can be less than the
bulk spin gap, and hence the ground state remains a singlet and the
spin-gap is preserved. All magnetic response functions will
therefore be exponentially small at low $T$.

At the bulk critical point, we need to examine the scaling
dimensions of the perturbations in $\mathcal{S}_{\rm imp}$. In the
$\epsilon=4-D$ expansion, simple power-counting shows that both
$\lambda_{1,2}$ are irrelevant. Consequently, we do not expect a
Curie spin susceptibility, but a suppressed response determined by
the scaling dimensions of $\lambda_{1,2}$. In the $D=2+\epsilon$
expansion,\cite{qimp2} we work with the constraint $|z_\alpha|^2 =
1$, and hence the $\lambda_2$ term disappears. The $\lambda_1$ term
is marginal at the tree level, and a more complete renormalization
group analysis is needed to understand the flow of $\lambda$; this,
however, is beyond the scope of the present paper.

Finally, for $s \ll 0$, in the magnetically ordered phase with
collinear order, we can carry out an analysis of the magnetic
properties as in Section~\ref{sec:neel}. We apply the ansatz in
Eq.~(\ref{zpsi}) to $\mathcal{S}_b+\mathcal{S}_{\rm imp}$ above, and
compute all physical properties as a perturbation series in $u_0$,
$v_{1,2}$ and $\lambda_{1,2}$. The result shows that a spatial
dependence of the non-collinear order and the uniform
(ferromagnetic) magnetization is induced near the impurity by
$\lambda_{1,2}$. The uniform magnetization requires the
`particle-hole' symmetry breaking term $\lambda_1$, and is non-zero
even in the absence of an applied magnetic field. However, unlike
the situation in Section~\ref{sec:neel}, this ferromagnetic moment
is not quantized, and takes a value dependent upon the bare values
of the couplings: this is because spin rotation symmetry is
completely broken in the non-collinear ordered phase.\cite{rmp}

\section{Conclusions}
\label{sec:conc}

Let us summarize the basic physical characteristics of the impurity
response of the various spin liquid states considered in this paper:

{\em (i) N\'eel-VBS transition:} The impurity characteristics across
this transition are qualitatively very similar to those found for
impurities in confining antiferromagnets,
\cite{science,qimplong,qimp2} although the underlying field theory
is quite different, as are all the exponents and scaling functions.
The spin-gap phase has the conventional Curie response of a local
moment at the impurity; for the confining antiferromagnets this
local moment is induced by confinement physics, while for the
N\'eel-VBS transition it is induced by the logarithmic `Coulomb'
interaction associated with the U(1) gauge field. The N\'eel-VBS
quantum critical point has an anomalous Curie response, with a Curie
constant $\mathcal{C}$ specified in Eq.~(\ref{chi-tot}) to leading
order in the $\epsilon=4-D$ expansion. The response of the N\'{e}el
phase is as found in the earlier case: a spatial varying staggered
and uniform moment is found near the impurity, with the moment in
the latter quantized as in Eq.~(\ref{magtotal}). The Knight shift
has both staggered and uniform components, with its behavior in the
vicinity of the impurity specified by the impurity exponents
determined in Section~\ref{sec:imp}. At the critical point, the
induced staggered moment is much stronger than the uniform
magnetization component and thus represents the dominating response
to an external field.

{\em (ii) Staggered flux spin liquid:} This spin liquid is
generically critical, and so has an anomalous Curie response; the
Curie constant is given, to the leading order in the $1/N_f$
expansion, by Eq.~(\ref{sfc}). The Knight shift now has only a
uniform component, but no staggered component in the scaling limit.
The absence of any staggered Knight shift is a defining
characteristic of the sF spin liquid, and is a consequence of its
large emergent symmetry.\cite{hermele2} There is a large number of
competing order parameters, and the leading impurity action
$\mathcal{S}_{\rm imp}$ has no natural way of choosing among them,
leading to a response which is restricted to the ferromagnetic
moment alone. The behavior of the ferromagnetic Knight shift upon
approaching the impurity is specified by the impurity exponent
(\ref{Delta-imp-SF}) found in Section~\ref{sec:imp2}; to the leading
order in $1/N_f$, the Fourier transform of the ferromagnetic Knight
shift is computed in Sect.\ \ref{sec:sfuni}.

{\em (iii) Spinon Fermi surface:} This case has a finite magnetic
response as $T\rightarrow 0$, unlike the divergent Curie
susceptibility of the sF phase. The response decays with a power law
away from the impurity, with the dominant response being at the
$2k_F$ wavevector.

{\em (iv) $Z_2$ spin liquid:} The spin liquid itself has an
exponentially suppressed impurity response, because the spin gap is
preserved in the presence of the impurity. The proximate ordered
non-collinear state has a response similar to that of the N\'eel
state, but with no quantization of the total magnetization.

Apart from their applications to spin liquids, the results in
Section~\ref{sec:sf} also have a direct application to the physics
of charged impurities in two-dimensional graphene. It is well known
that the low energy electronic excitations in graphene are described
by 2 species of Dirac fermions. There is no fluctuating gauge field
$A_\mu$ as in Eq.~(\ref{diracaction}). However in the presence of
charged impurity, the {\em three-dimensional\/} Coulomb potential
the electrons experience has the form $C/r$, where $C$ is a
constant. Interestingly, this is exactly the form of the static
potential found in Section~\ref{sec:sf} in the presence of an
impurity, where $\langle A_\tau \rangle \sim 1/r$. Thus, as long as
we ignore quantum-electrodynamic loop corrections, the results of
Section~\ref{sec:sf} apply also to graphene; specifically, in
Fig.~\ref{fig:psi}, (a) and (b) apply to graphene while (c)-(h) do
not. One of our important results for this system was
Eq.~(\ref{psi0}): that the induced charge density due to the Coulomb
potential is a delta function, with $Q$ a non-trivial universal
function of $C$. This results therefore applies also to graphene. It
disagrees with the earlier result of Ref.~\onlinecite{mele}, which
found a $1/r^2$ decay in the induced charge density. We believe
their results suffers from a cavalier treatment of the ultraviolet
cutoff, which does not preserve gauge invariance.

We now discuss possibilities for future theoretical work.

For the sF and ${\rm CP}^{N-1}$ spin liquids, an alternative $1/N_f$
(or respectively $1/N$) expansion could give a useful complementary
picture. We have used an expansion here in which the $N_f \to
\infty$ was taken at fixed $Q$. However, it is also possible to
study the limit $N_f \to \infty$ limit at fixed $Q/N_f$. The
saddle-point equations are straightforward to obtain for this case,
but much more difficult to solve: they have the advantage of
including the spinon-impurity `Coulomb' interaction at all orders
already at $N_f=\infty$, and so allow for spinon states below, or at
the edge of, the bulk spinon continuum. We will report on the
results of such an analysis in future work.

It would also be useful to have a more complete understanding of the
spinon Fermi surface case, by extending the work of
Refs.~\onlinecite{aim,km} to determine the gauge-invariant response
functions near a single impurity.

Finally, we comment on the experimental implications of our work.

For the cuprates, NMR experiments\cite{bobroff,ouazi} show a large
Knight shift response at the N\'eel wavevector in the vicinity of
the impurity, and strong temperature-dependent enhancement of such
correlations. These features are consistent with the N\'eel-VBS
transition considered in Section~\ref{sec:cpn}, but also with the
transition in dimerized antiferromagnets considered
earlier.\cite{science,qimplong,qimp2} The absence of a staggered
response for the sF case is potentially a serious deficiency of the
model of Section~\ref{sec:sf}. It remains to be seen if corrections
to scaling (such as those in Eq.~(\ref{Vpert})) can remedy the
situation.

For the organic Mott insulator $\kappa$-(ET)$_2$Cu$_2$(CN)$_3$, the
NMR Knight shift \cite{shimizu} has an appreciable $T$ independent
component at low $T$. This can potentially be fit either by the
spinon Fermi surface, or by a weakly magnetically ordered state.

\acknowledgments

We thank A.~Castro Neto, M.~Hermele, Y.~B.~Kim, T.~Senthil, and
M.~Vojta for useful discussions. This research was supported by the
National Science Foundation under grant DMR-0537077. AK was
supported by the grant KO2335/1-1 under the Heisenberg Program of
Deutsche Forschungsgemeinschaft.

\end{document}